\documentstyle[12pt,epsfig]{article}
\begin{document}
\begin{quote}
\raggedleft
hep-ph/yymmxxx \\
LPM-96-12 \\
CERN-TH/96-64 \\
March 1996 
\end{quote}
\begin{center}
{\bf \Large Radiative Contributions to TGC in the MSSM~\footnote{
To appear in {\it Proceedings of the Physics with $e^+e^-$ Linear 
Colliders Workshop}, 
Annecy -- Gran Sasso -- Hamburg 1995, ed. P. Zerwas} \\}
\vspace*{0.3cm}
A. Arhrib $^a$, J.-L. Kneur $^{b,}$\footnote{On leave
from U.R.A. 768 du C.N.R.S., F34095 Montpellier France.}
and G. Moultaka $^a$ \\
\vspace*{5mm}
$^a$ Physique Math\'ematique et Th\'eorique, U.R.A. du 
CNRS N$^o$ 040768, \\ 
Universit\'e Montpellier II, F-34095 Montpellier France\\
\vspace*{2mm}
$^b$ Theory Division, CERN, \\
CH-1211 Geneva 23, Switzerland \\
\vspace{1cm}
\abstract{We give a brief account of recent calculations of radiative
contributions to the Triple Gauge Couplings (TGC) from the Minimal
Supersymmetric Standard Model (MSSM), at a 500 GeV $e^+e^-$ collider. 
Our results indicate that, although these MSSM virtual 
contributions indeed are 
of the order of the expected accuracy on TGC measurements, 
the generally neglected box contributions to TGC also are likely to be
relevant at such high energies. }  

\end{center}

\newpage 
\setcounter{footnote}{0} 
\begin{center}
{\bf \Large Radiative Contributions to TGC in the MSSM \\}
\vspace*{0.3cm}
A. Arhrib $^a$, J.-L. Kneur $^{b,}$\footnote{On leave
from U.R.A. 768 du C.N.R.S., F34095 Montpellier France.}
and G. Moultaka $^a$ \\
\vspace*{5mm}
$^a$ LPM, U.R.A. du CNRS N$^o$ 040768, Univ. Montpellier II, France\\ 
\vspace*{2mm}
$^b$ Theory Division, CERN, 
Geneva, Switzerland \\
\end{center}
\baselineskip=18pt
\vspace*{1cm}
%
\newcommand{\beq}{\begin{equation}}
\newcommand{\eeq}{\end{equation}}
\newcommand{\bea}{\begin{eqnarray}}
\newcommand{\eea}{\end{eqnarray}}
%
\section{Radiative Contributions to TGC}
With an expected accuracy of O(10$^{-3}$) or better~\cite{nlctgc}, 
the measurement of
Triple Gauge Couplings (TGC) at a 0.5--2 TeV $e^+e^-$ collider 
will truly constitute a {\it precision} experiment. Therefore, apart
from testing possible departures from the SM $WW\gamma$, $WWZ$ vertices
at tree-level, with an accuracy improved by more than an order of
magnitude with respect to LEP2~\cite{lep2tgc}, it
is legitimate to expect detectable loop-level TGC 
contributions from different models of
New Physics. 
Actually, one-loop TGC contributions are certainly present
in any renormalizable model, 
but such virtual effects are suppressed by a factor of
$(g^2/16\pi^2) \simeq $ 2.7 10$^{-3}$. Moreover, further 
suppression is expected
from the decoupling properties of heavy particles in most 
renormalizable
models.  
For instance, SM one-loop TGC 
predictions are known~\cite{smtgc,pinch} and give, at $\sqrt s = 
$ 190 GeV~\cite{lep2tgc} (
$\sqrt s = $ 500 GeV):
$\Delta \kappa_\gamma \simeq $  4.1--5.7 $\times$ 10$^{-3}$
(5.5--($-$5.4) $\times$ 10$^{-4}$), for $M_{Higgs} =$ 0.065--1 TeV and
$m_{top}=$ 175 GeV
(with $\Delta \kappa_Z$ of the same order and 
even smaller contributions to $\lambda_{\gamma ,Z}$).  
The smallness of those SM contributions at 500 GeV is due to
their fast decreasing at high energies, $q^2 >> M^2_{W,Z},m^2_{top}$,
in accordance with the good unitarity behaviour, while even 
for a heavy Higgs
its effects in $\Delta \kappa_{\gamma ,Z}$ 
are screened, giving non-decoupling but small (constant) 
contributions for $M_{Higgs} \to \infty $. 

A more interesting situation is thus expected  
if, for example, some particles in the loops have a stronger 
non-decoupling behavior
(e.g due to Yukawa type couplings) 
and/or are close to their production threshold. The latter is 
likely to be
the case
in the Minimal Supersymmetric Standard Model (MSSM)~\cite{mssm} 
since, as is well known, the resolution of the hierarchy problem 
requires the 
spectrum of supersymmetric partners to appear at O(1 TeV) or below. 
A study of MSSM one-loop contributions to 
TGC~\cite{mssmtgc1,mssmtgc2} 
can therefore
provide a complementary information on a range of 
MSSM parameter values which may not be available from
direct particle production.

\section{Extracting TGC from Loops}
In momentum space the vertex issued from the C-, P-conserving 
part
 of
the general TGC effective Lagrangian~\cite{tgcgen} reads 
($ V \equiv \lambda , Z$) 
\bea
\label{vertex}
\Gamma^V_{\mu \alpha \beta} = i g_{VWW} \{f_V
[2g_{\alpha\beta} \Delta_\mu +4(g_{\alpha\mu} Q_\beta-g_{\beta\mu}
Q_\alpha )] \\ \nonumber
+2\Delta\kappa^\prime_V(g_{\alpha\mu}Q_\beta -g_{\beta\mu}Q_\alpha)
+4{\Delta Q_V \over M^2_W} \Delta_\mu(Q_\alpha Q_\beta -
g_{\alpha \beta}{Q^2\over 2}) \},
\eea
where
$2Q_\mu$, $(\Delta-Q)_\alpha$, and $-(\Delta+Q)_\beta$ designate the
four-momenta and Lorentz
indices of the {\it incoming} $\gamma$ (or $Z$),  $W^+$, and
$W^-$, respectively, and the ($q^2$-dependent) 
coefficients in (\ref{vertex}) are related   
to the more conventional TGC parameters~\cite{tgcgen,lep2tgc} as \\ 
$ \Delta\kappa^\prime_V \equiv \kappa_V -1 +\lambda_V \;
=\Delta \kappa_V
+\lambda_V $; 
$ \Delta Q_V \equiv -2 \lambda_V $.
 
Naively, TGC are obtained by
summing all MSSM contributions to the appropriate parts in eq. 
(\ref{vertex})
from vertex loops with entering $\gamma$ (or $Z$) and
outgoing  $W^+$, $W^-$.
But, as is well known, the vertex graphs with virtual gauge bosons
need to be combined with parts of box graphs for the full process,
$e^+e^- \to W^+W^-$, to form a gauge-invariant contribution.
This is most conveniently done 
by the pinch technique~\cite{pinch}, i.e `pinching' in an appropriate
manner the irrelevant propagator
lines from boxes, 
which preserves all the
 well-behaved features and properties expected from radiative
corrections
(Ward identities, good unitarity behaviour, infra-red
finiteness, etc).
The resulting combinations define purely $s$-dependent
--and in that sense universal--  TGC since, 
by definition, $t$ and $u$-dependent
box contributions are left over in this procedure.
In what follows we have evaluated the complete set of $s$-dependent
TGC, using the pinch technique for the relevant graphs. 
The sfermions, gauginos, and Higgses contribute separately to
vertex graphs 
at the one-loop level, and we illustrate in fig. 1 those three 
contributions to $ \Delta\kappa_\gamma $ 
for a representative choice of 
MSSM parameters~\footnote{For illustration we choose values which 
give somewhat maximal 
effects.} (contributions to $\Delta\kappa_Z $ are of the same 
order~\cite{mssmtgc2}).
As a general remark, we note that the total contributions from 
the three sectors 
are comfortably of the order of the expected accuracy, but most
particles give their maximal
contributions when their masses are slightly above
their direct production thresholds, showing a  
decoupling behavior for larger masses, as expected. An exception 
occurs 
for the sfermions, whose largest (and indeed dominant) 
contributions are
obtained for a large mass splitting among up and down 
components (a
situation where one does not expect any decoupling behavior, 
and in fact $\Delta\kappa_V$ tend to a constant in that case). 

Now it is not clear whether 
the remnant (u,s)- and (t,s)-dependent boxes 
--which also form a gauge-invariant set by 
themselves--
do not contribute a substantial part
of the full radiative corrections to TGC, especially at 
such high energies. 
Even if the above concept of
a gauge-invariant, universal quantity defined from 
the pinch technique could be
theoretically useful, one practical problem 
(as far as TGC measurements are
concerned) is that there are no planned experimental procedure 
to distinguish
among ``universal" and ``remnant" TGCs: clearly 
{\it anything}
contributing at the loop-level to the coefficients in 
eq.(\ref{vertex})
will be extracted from the data (provided, of course, 
that it gives large
enough contribution to be detected).
We thus illustrate as well in fig. 2 some partial (but
gauge-invariant) 
box contributions,
with internal sleptons and gauginos. 
Although partial, this indicates what relative amount of 
universal versus ``remnant box" contributions may be 
expected at those
energies.

\begin{figure}[htb]
 \mbox{\epsfig{file=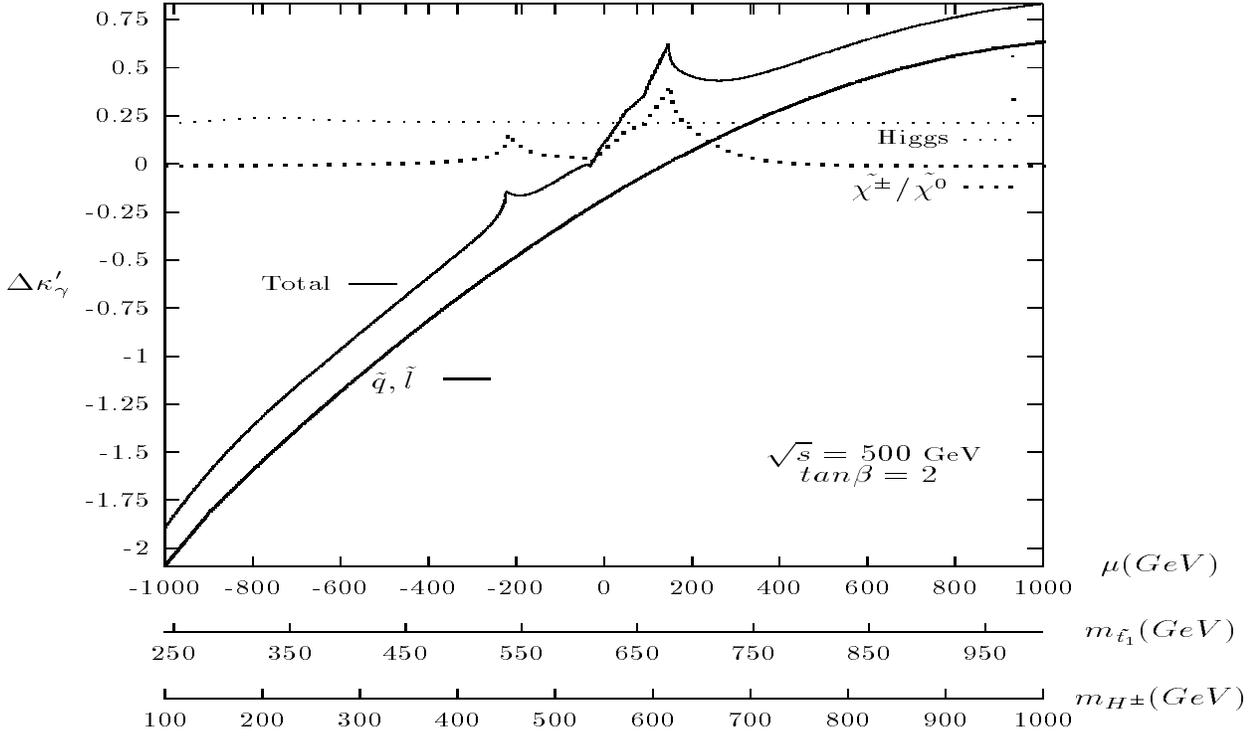,height=8cm,width=10cm,
bbllx=100,bblly=400,
bburx=350,bbury=700}}
\vspace*{1.5cm}
\caption{\label{fig1} 
s-dependent $\Delta\kappa^\prime _\gamma $ 
contributions (in units of $g^2/16\pi^2$) from 
Higgses, sfermions and gauginos. Other parameters fixed to 
$m_{\tilde{t}_1}=m_{\tilde{t}_2}=m_{\tilde{U}_1}=m_{\tilde{U}_2}
=m_{\tilde{{
\it l}}_1}=m_{\tilde{{\it l}}_2}$; 
$m_{\tilde{t}_1} + m_{\tilde{\nu}_L}=$ 1.245 TeV and 
$m_{\tilde{t}_1} + m_{
\tilde{D}_{1,2}}=$ 1.47 TeV;  $M=$190 GeV, $M'=$70 GeV.} 
\end{figure}
\begin{figure}[htb]
 \mbox{\epsfig{file=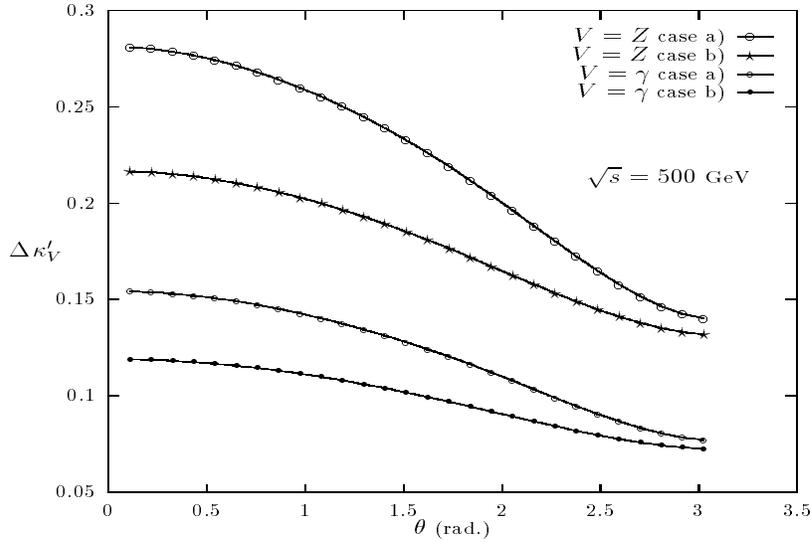,height=8cm,width=10cm,
bbllx=70,bblly=400,
bburx=400,bbury=750}}
\caption{\label{fig2} 
t-dependent non-pinch box contributions 
(in units of $g^2/16\pi^2$) with one
chargino (resp. neutralino),
two sneutrinos (resp. selectrons) and one selectron (resp.
 sneutrino) versus the $W^-$ production angle $\theta$.  
$m_{\tilde{e}_1}=m_{\tilde{\nu}_L}=$ 260 GeV, zero left-right
mixing angle;
case a) $M=\mu=$150 GeV, $M'=$100 GeV, $\tan\beta =15$;
case b)  $M=M'=\mu=$ 250 GeV, $\tan\beta =2$.}
\end{figure}

\end{document}